## Estimating Open Access Mandate Effectiveness: The MELIBEA Score

**Philippe Vincent-Lamarre[1,4], Jade Boivin[1], Yassine Gargouri[1],**

**Vincent Larivière[2], Stevan Harnad[1,3]**

[1]Université du Québec à Montréal, [2]Université de Montréal, [3]University of Southampton, [4]Université d'Ottawa

**Abstract:** MELIBEA is a directory of institutional open-access policies for research output that uses a composite formula with eight weighted conditions to estimate the "strength" of Open Access mandates (registered in ROARMAP). We analyzed total Web of Science-(WoS)-indexed publication output in years 2011-2013 for 67 institutions where OA was mandated in order to estimate the mandates' *effectiveness*: How well did the MELIBEA score and its individual conditions predict *what percentage* of the WoS-indexed articles is actually deposited in each institution's OA repository, and *when*. We found a small but significant positive correlation (0.18) between the MELIBEA "strength" score and deposit percentage. For three of the eight MELIBEA conditions (deposit timing, internal use, and opt-outs), one value of each was strongly associated with deposit percentage or latency (1: immediate deposit required; 2: deposit required for performance evaluation; 3: unconditional opt-out allowed for the OA requirement but no opt-out for deposit requirement). When we updated the initial values and weights of the MELIBEA formula to reflect the empirical association we had found, the score's predictive power for mandate effectiveness doubled (.36). There are not yet enough OA mandates to test further mandate conditions that might contribute to mandate effectiveness, but the present findings already suggest that it would be productive for existing and future mandates to adopt the three identified conditions so as to maximize their effectiveness, and thereby the growth of OA.

## Introduction

The Open Access (OA) movement (BOAI 2002) arose as a result of two concurrent developments: (1) the "serials crisis," which made research journals increasingly unaffordable, hence inaccessible, to researchers' institutions (Okerson & O'Donnell 1995; Miller et al 2010), even the richest ones (Harvard University Library 2012) and (2) the advent of the online medium, which made it possible in principle to make all research journal articles freely accessible to all users online.

The primary target content of the OA movement is refereed research journal articles. That is the only kind of item analyzed in this study. Researchers can provide OA to their journal articles in two different ways – by publishing in an OA journal ("Gold OA," often for a publication fee) or by publishing in a subscription journal and, in addition, *self-archiving* the final, peer-reviewed draft online ("Green OA") (Harnad et al 2004). The present analysis of institutional OA policies is based exclusively on articles published in non-OA journals. Those subscription journal articles are the primary targets of *institutional OA self-archiving mandates* (because mandate effectiveness is dependent on author compliance and constrained by subscription publisher OA embargoes), whereas articles in OA journals are OA already (whether or not they are self-archived). Our dependent variable in this study is *the percentage of each mandating institution's articles published in non-OA journals that are deposited in the author's institutional repository* (as well as the deposit's timing and OA status).

OA also comes in two degrees: "Gratis" OA is free online access. "Libre" OA is free online access plus certain re-use rights (Suber 2008a). We will only be considering Gratis OA in this study, because (1) Gratis OA is a prerequisite for Libre OA, (2) it faces fewer publisher restrictions (e.g., OA embargoes), and (3) it is the most urgently needed by researchers. We will also only be considering Green OA (self-archiving), because (a) the majority of OA to date is Green, (b) it does not entail any payment of OA publishing fees, (c) it can be provided by researchers themselves, and, as will be explained below, (d) providing Green OA can be mandated by researchers' institutions and funders (without having to pay any extra publisher Gold OA fees).

One might have expected that because of the many advantages provided by OA (Hitchcock 2013) – *viz*, free access for all users, enhanced research uptake and impact, relief from the serials crisis, and the speed and power of the online medium – researchers would all have hastened to make their papers (Green) OA ever since it became possible. But the growth of OA – which has been possible since even before the birth of the Web in 1989 (Berners-Lee 1989) -- has actually been surprisingly slow: For example, in 2009, already two decades after the Web began, Björk et al estimated the percentage of OA to be only 20.4% (the majority of it Green). In 2013 Archambault announced close to 50% OA, but Chen (2014) found only 37.8% and Khabsa & Giles (2014) even less (24%). All these studies reported great variation across fields and *none took timing adequately into account* (publication date vs. OA date): Björk et al (2014) pointed out that 62% of journals (from the top 100 journal publishers indexed by SCOPUS) endorse immediate Green OA self-archiving by their authors, 4% impose a 6-month embargo, and 13% impose a 12-month embargo; so at least 79% of articles published in any recent year could already have been OA within 12 months of their date of publication via Green OA alone, 62% of them immediately, if authors were actually providing it.

There are many reasons why researchers have been so slow to provide Green OA even though they themselves would be its biggest potential beneficiaries (Spezi et al 2013; Taylor & Francis 2015; Wallace 2011). The three principal reasons are that (i) researchers are unsure whether they have the legal right to self-archive, (ii) they fear that it might put their paper's acceptance for publication at risk and (iii) they believe that self-archiving may be a lot of work (Harnad 2006). All these concerns are demonstrably groundless (Pinfield 2001; Swan & Brown 2005; Carr et al 2005, 2007; ETH Zurich FAQ 2015; SHERPA/Romeo 2015) but it has become increasingly clear with time that merely pointing out how and why they are groundless is still not enough to induce authors to go ahead and self-archive of their own accord. Researchers' funders and institutions worldwide are now beginning to realize that they need to extend their existing "publish or perish" mandates so as to make it *mandatory* to provide OA to researchers' publications, not only for the benefit of (i) the researchers themselves and of (ii) research progress, but also to maximize the (iii) return to the tax-paying public on its investment in funding the research.

Funders and institutions are accordingly beginning to adopt OA policies: Starting with NIH in the US and the Wellcome Trust in the UK -- soon followed by the Research Funding Councils UK, the European Commission, and most recently President Obama's Directive to all the major US federal funding agencies -- research funders the world over are beginning to mandate OA (ROARMAP 2015). In addition, research institutions, the providers of the research, are doing likewise, with Harvard, MIT, University College London and ETH Zurich among the vanguard, adopting OA mandates of their own that require all their journal article output, across all disciplines, funded and unfunded, to be deposited in their institutional OA repositories.

These first OA mandates, however, differ widely, both in their specific requirements and in their resultant success in generating OA. Some mandates generate deposit rates of over 80%, whereas others are doing no better than the global baseline for spontaneous (un-mandated) self-archiving (Gargouri et al 2013). As mandate adoption grows worldwide, it is therefore important to analyze the existing mandates to determine which mandate conditions are important to making a mandate effective (Swan et al 2015).

We have accordingly analyzed the institutional policies indexed by MELIBEA, a directory of institutional open-access policies for research output that classifies OA mandates in terms of their specific conditions as well as providing a score for overall mandate *strength*. This score is based on a composite formula initialized with a-priori values and weights for eight conditions that MELIBEA interprets as reflecting policy strength (see Figure 3 and Table 1). The purpose of the present study was (i) to test the predictive power of the overall MELIBEA score for mandate *effectiveness*, as reflected in deposit rate and deposit timing, (ii) to test individually the association of each of the conditions with deposit rate and timing, and (iii) to transform the initial values and weights of the MELIBEA formula to reflect the empirical findings of (ii).

What we found was a small but significant positive correlation between (i) the original MELIBEA score for policy strength and (ii) deposit rate. We also found that for three of the eight MELIBEA conditions (1: deposit timing, 2: internal use, and 3: opt-outs), one of the value options for each condition was most strongly correlated with deposit rate and latency (1: immediate deposit required, 2: deposit required for performance evaluation, 3: unconditional opt-out allowed for the OA requirement but no opt-out for deposit requirement). When we updated the initial values and weights of the original MELIBEA formula to reflect this correlation, the new score's predictive power for deposit rate was doubled. This suggests that it would be useful to adopt OA mandates with these conditions in order to maximize their effectiveness, and thereby maximize the growth of OA.

## Databases Used

The data for our analysis were drawn from several databases. The *ROAR Registry of OA Repositories* provided a database of all open access repositories (ROAR, http://roar.eprints.org). The *ROARMAP Registry of OA Repository Mandates and Archiving Policies* provided the subset of the ROAR repositories that had an OA mandate (ROARMAP, http://roarmap.eprints.org). An in-house version of the *Thomson Reuters Web of Science* (WoS) database hosted at the *Observatoire des sciences et des technologies* (OST-UQAM) provided the bibliographic metadata for all (WoS-indexed) articles published in years 2011-2013 by any author affiliated with the ROARMAP subset of institutions with OA mandates that had been adopted by 2011. The *MELIBEA Directory and Estimator of OA Policies* provided a classification of the OA mandates in terms of their specific conditions (assigning a numerical value to each option for each condition) plus an overall score based on a weighted combination of eight of those conditions as an estimate of policy strength (MELIBEA (http://accesoabierto.net/politicas) (Table 1).

WoS indexes the 12,000 most cited peer-reviewed academic journals in each research area. (This corresponds to slightly more than 11% of the 105,000 peer-reviewed journals listed in Ulrich's Global Serials Directory, http://ulrichsweb.com.)  The bibliographic records (authors, title, journal, volume, issue, date) for all the WoS-indexed articles for each of the three years (2011-2013) for each of the target institutions were retrieved from the WoS database. The bibliographic records (authors, title, journal, volume, issue, date) for all the deposited full-texts for each of the three years (2011-2013) for each of the target institutions were retrieved from the repository database. The percentage deposit was determined by the percentage of WoS records that matched the repository records. (Note that this is not the percentage of all institutional output, only of WoS-indexed output; nor is it the percentage of all WoS-indexed output that is OA somewhere on the web, but only the percentage that is deposited in the institutional repository, in compliance with the institution's OA policy.)

We also used the *Webometrics Ranking Web of Universities*, which estimates institutions' "excellence" based on how many of their published articles are among the most cited 10% of articles (the lower the score the higher the rank). (http://webometrics.info). A disjoint set 19 institutions at the very bottom of the distribution (i.e., those with Webometrics ranks beyond 5000th; see Figure 1) was excluded from our sample as outliers. To balance representativeness and sample size and minimize sampling bias, we further excluded any institution that had fewer than 30 publications for any of the three years in our study. The number of institutions resulting after applying each of these selection criteria is shown in Table 1.

| | RoarMap (R) | Melibea (M) | R∩M | Sample size(R∩M) | Total |
|---|---|---|---|---|---|
| 2011 | 141 | - | 120 | 51 | 47 |
| 2012 | 162 | - | 140 | 62 | 57 |
| 2013 | 194 | - | 164 | 71 | 65 |
| 2011-3 | 194 | 426 | 164 | 77 | 67 |

**Table 1:** Total number of institutional policies indexed by ROARMAP (R), MELIBEA (M) and their shared intersection (R∩M), followed by the minimal sample size and the total number of institutions left after applying our selection criteria. Policy onset dates were based on ROARMAP.

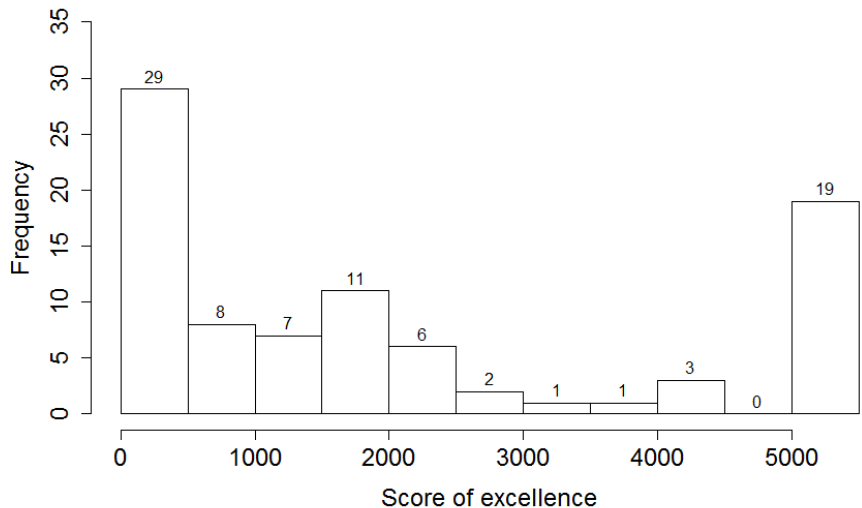

Figure 1: Distribution of mandated institutions by their Webometrics ranks for "excellence" (citedness), (Note: Higher score means lower rank. The 19 bottom-ranked outliers were excluded from study.)

**Deposit Rate and Deposit Latency**

Each of the 67 institutional repositories was crawled to determine *what percentage* of its WoS-indexed articles had been deposited (*deposit rate*), and *when* they were deposited

(*deposit latency*, i.e., the delay between (1) the publication date according to the WoS Access date and (2) the repository deposit date) for publication years 2011, 2012 and 2013 (crawled in April 2014). Deposits could be of two kinds: *Open Access (OA)* or *Restricted Access (RA)*, to comply with publisher OA embargoes. We subtracted the publication date from the deposit date, so negative values mean an article was deposited before publication and positive values mean it was deposited after publication (Figure 2).

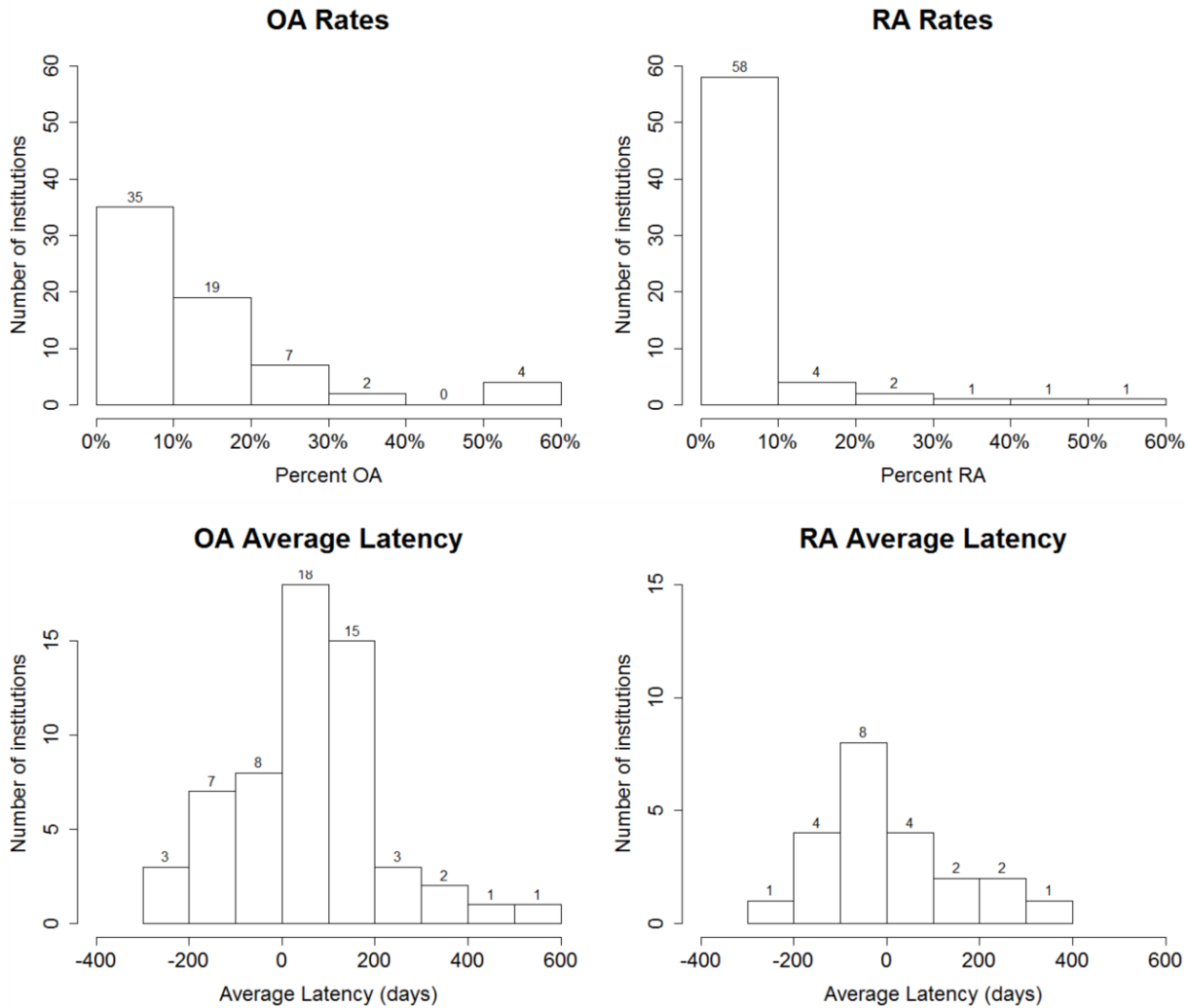

Figure 2: Distribution of institutions' average deposit rates (percentage of WoS articles deposited) and deposit latencies (in days, relative to publication date) for Open Access (OA) and Restricted Access (RA) deposits. The peak for RA deposit latency is negative (i.e., before publication) and the peak for OA deposit latency is positive (i.e., after publication).

**MELIBEA Score for Mandate Strength**

We used each mandating institution's overall MELIBEA score for policy strength as of April 6 2014. This score is a combination of eight specific individual conditions (re-numbered here, with each condition weighted by a percentage weight, summing to 100%) plus the respective options for each condition (each option weighted by a numerical value) (Figure 3 and Table 2) for each OA policy for each mandating institution. We tested how well the MELIBEA strength formula consisting of these 8 weighted conditions and their weighted option values jointly predicted deposit rate and deposit latency (i.e., policy effectiveness). Then we analyzed each individual condition and its options separately with Analyses of Variance (ANOVA) to determine how much each individual option contributed to predicting deposit rate or latency.

$$40\%(C1) + 10\%(C2) + 5\%(C3) + 10\%(C4)\ 5\%\ (C5) +\ 10\%(C6) +\ 10\%(C7) +\ 10\%(C8)$$

Figure 3: The eight mandate conditions (renamed C1-C8 here) and their initial percentage weights in MELIBEA's formula for policy strength. (See Table 2 below for a description of each condition and its respective options and value weights.)

| C1: Mandate or Request (40%)[1] | | C5 : Embargo Length(10%)[2] | |
|---|---|---|---|
| **Value** | | -2 | 12 months after publication |
| 1 | Request | 0.5 | 6 months after publication |
| 2 | Requirement | -2 | More than 12 months after publication |
| **C2 : Opt-out (10%)** | | -2 | Unspecified ('after period stipulated by the publisher') |
| 1 | No[2] | C6 : Copyright (10%)[2,3] | |
| **2** | **No deposit opt-out but unconditional OA opt-out** | 2 | Require blanket copyright reservation, no opt-out |
| **2** | **No deposit opt-out but conditional OA opt-out** | 1 | Allow opt-out from copyright reservation case by case |
| -1 | Deposit opt-out and unconditional OA opt-out[2] | 2 | Retain copyright and license specific rights to publisher |
| **C3 : Version (5%)[2]** | | 2 | Authors retain the non-exclusive rights of explotation to allow self-archiving |
| 0.8 | Deposit author's version | 1 | Authors should retain copyright whenever is possible |
| 0.8 | Deposit publisher's version | 1 | Any publishing or copyright agreements concerning articles have to comply with the OA policy |
| 0 | Unspecified | -2 | No copyright reservation |
| 0.4 | Unrefereed preprint | C7 : Internal use (10%) | |
| **C4 : Deposit Timing (10%)** | | **2** | **Yes** |
| 0.5 | As soon as possible[2] | **0** | **No** |
| **2** | **At time of acceptance** | C8 : Theses (10%)[4] | |
| **1.5** | **At time of publication** | 2 | Yes |
| **0** | **Unspecified** | 0 | No |

[1]Only mandatory (required) policies considered in this analysis, not recommended policies

[2]Not enough instances

[3]Inconsistent options

[4]Only journal article conditions relevant

Table 2: The eight MELIBEA conditions (re-numbered) together with the initial values assigned by MELIBEA for each option for each condition in the formula for the mandate strength score, Figure 3). Separate ANOVAS were done (see below) for the three **boldface** conditions and options *(2, 4, 7)*. The other conditions either did not have enough mandate instances (3, 5, 6) or were irrelevant to this study (1, 8).

We excluded from this study all OA policies that were not mandates (i.e., compliance was not required, only recommended) because it is already known that mandating (i.e., requiring) OA is more effective than just requesting, recommending or encouraging it (Suber 2008b, Xia et al 2012). Our specific interest here is in *which of the conditions of mandatory OA policies make a contribution to increasing deposit rate and/or decreasing deposit latency*. For the individual t-tests and ANOVAs testing the eight MELIBEA conditions and their respective options separately, three of the conditions (Version (C3), Embargo Length (C5), Copyright (C6)) did not have enough institutional mandate instances to be tested with an ANOVA; one condition (Theses (C8)) was irrelevant to our target content, which was only refereed journal articles; and, as noted, for C1 (Required or Recommended), our study concerned only the mandatory (required) policies, not recommended ones. We accordingly analyzed only conditions *C2 (Opt-Out), C4 (Deposit Timing)* and *C7 (Internal Use)* with their respective options:

*Condition C2 (Opt-Out):* Deposit and OA are both required and (2a) opt-out is not allowed from the deposit requirement but <u>conditional</u> opt-out is allowed from the OA requirement, on a case-by-case basis; (2b) opt-out is not allowed from the deposit requirement but <u>unconditional</u> opt-out is allowed from the OA requirement; (2c) unconditional opt-out is allowed from both the deposit and the OA requirements.

*Condition C4 (Deposit Timing):* Deposit is required (4a) at time of acceptance, (4b) at time of publication, or (4c) time unspecified.

*Condition C7 (Internal Use):* Deposit is (7a) required for internal use (usually research evaluation) or (7b) not required for internal use.

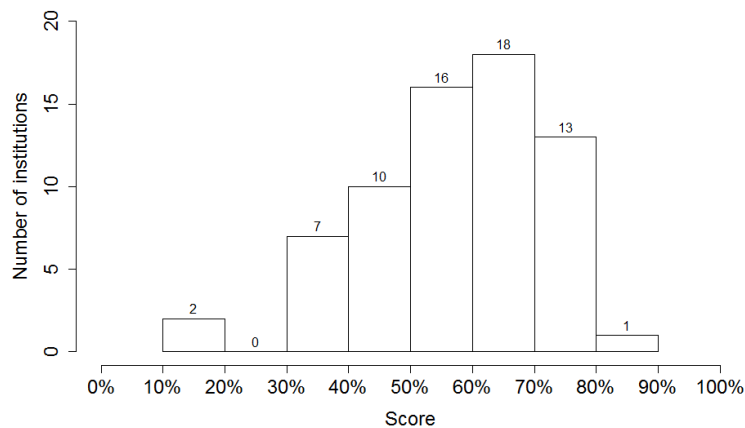

Figure 4: Distribution of MELIBEA scores for OA mandate strength

**Power of Initialized MELIBEA Score to Predict Deposit Rate and Latency**

To test the correlation between the MELIBEA score for mandate strength and deposit (1) rate and (2) latency, we analyzed OA and RA deposits separately as well as jointly (OA + RA). For deposit latency, which was normally distributed (its skewness fell within two Standard Errors of Skewness [SES]), we used the Pearson product moment correlation coefficient. For deposit rate, which was not normally distributed (Figure 2), we used permutation tests[1] (Edgington & Onghena, 2007). The individual mandate conditions were then tested using a combination of t-tests and one-way ANOVAs, with and without permutation testing.

The overall MELIBEA score is a weighted combination of the eight specific conditions of OA policies (Table 2), assigning a value to each of their options, and then combining them into the weighted formula in Figure 1 ($40\%(C_1) + 10\%(C_2) + 5\%(C_3) + 10\%(C_4) + 5\%(C_5) + 10\%(C_6) + 10\%(C_7) + 10\%(C_8)$).

MELIBEA initialized the overall score with these weightings on the 8 conditions as well as with their respective option values (on the basis of theory and what prior evidence was available). Each of these initialized values and weightings is now awaiting evidence-based validation of their power to predict the effectiveness of OA mandates (i.e., how many OA and RA deposits they generate, and how soon) so that either the values or the weightings can be updated to maximize their predictive power as the evidence base grows. At the end of this section we will illustrate how these initial values and weightings can be updated based on the present findings so as to make the MELIBEA score more predictive of deposit rate and latency.

Our analysis found that this initial overall MELIBEA score has a small significant positive correlation with deposit rate for OA deposits (Figure 5) but not for RA deposits (nor for OA + RA jointly). For deposit latency there is no correlation at all with the overall MELIBEA score.

---

[1] Permutation tests consist of random permutations of the values in a sample iterated a large number of times (e.g. 10,000) and computing the test statistic for each new sample. The *p* value is the probability that the test statistic of the original sample will occur within the distribution generated. Permutation testing does not presuppose homogeneity of variance or normality.

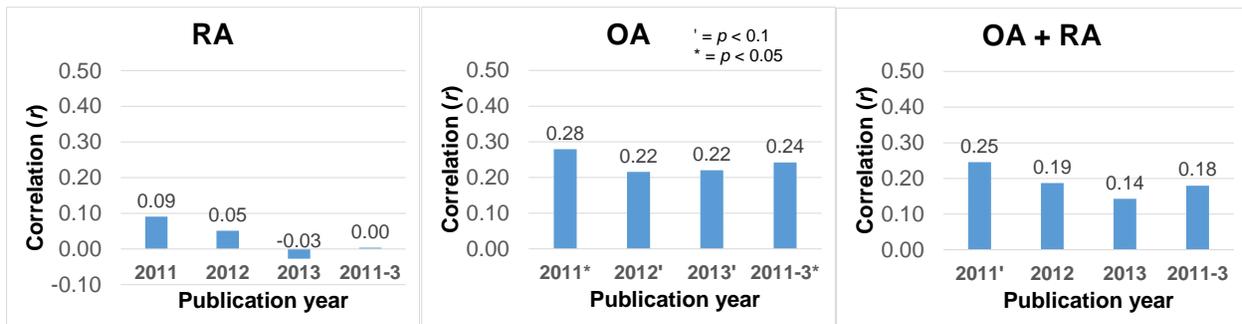

Figure 5: Correlation between MELIBEA mandate strength score and deposit rate, by publication year, for RA deposits, OA deposits and RA + OA deposits jointly

## Effect of Deposit-Timing Condition (C4) on Deposit Rate and Latency

To update and optimize the weightings on the specific OA policy conditions, we need to look at their predictive power individually. As noted, two of the eight conditions are not relevant for this study, and for three of the conditions the number of OA policies that have adopted them is so far still too small to test their effectiveness. We report here the results for only those three conditions (*Opt-Out (C2), Deposit Timing (C4)* and *Internal Use (C7))* that have reached testable sample sizes.

For OA and RA deposits combined, mandates that required deposit "*At time of acceptance*" had a significantly higher deposit rate than mandates that required deposit "*At time of publication*" or "*Unspecified.*" The same pattern is present for each of the three years, significant for year 2011 and almost significant for all three years combined (Figure 6). The effect for RA deposits alone and OA deposits likewise shows exactly the same pattern, for each of the three years, but without reaching statistical significance.

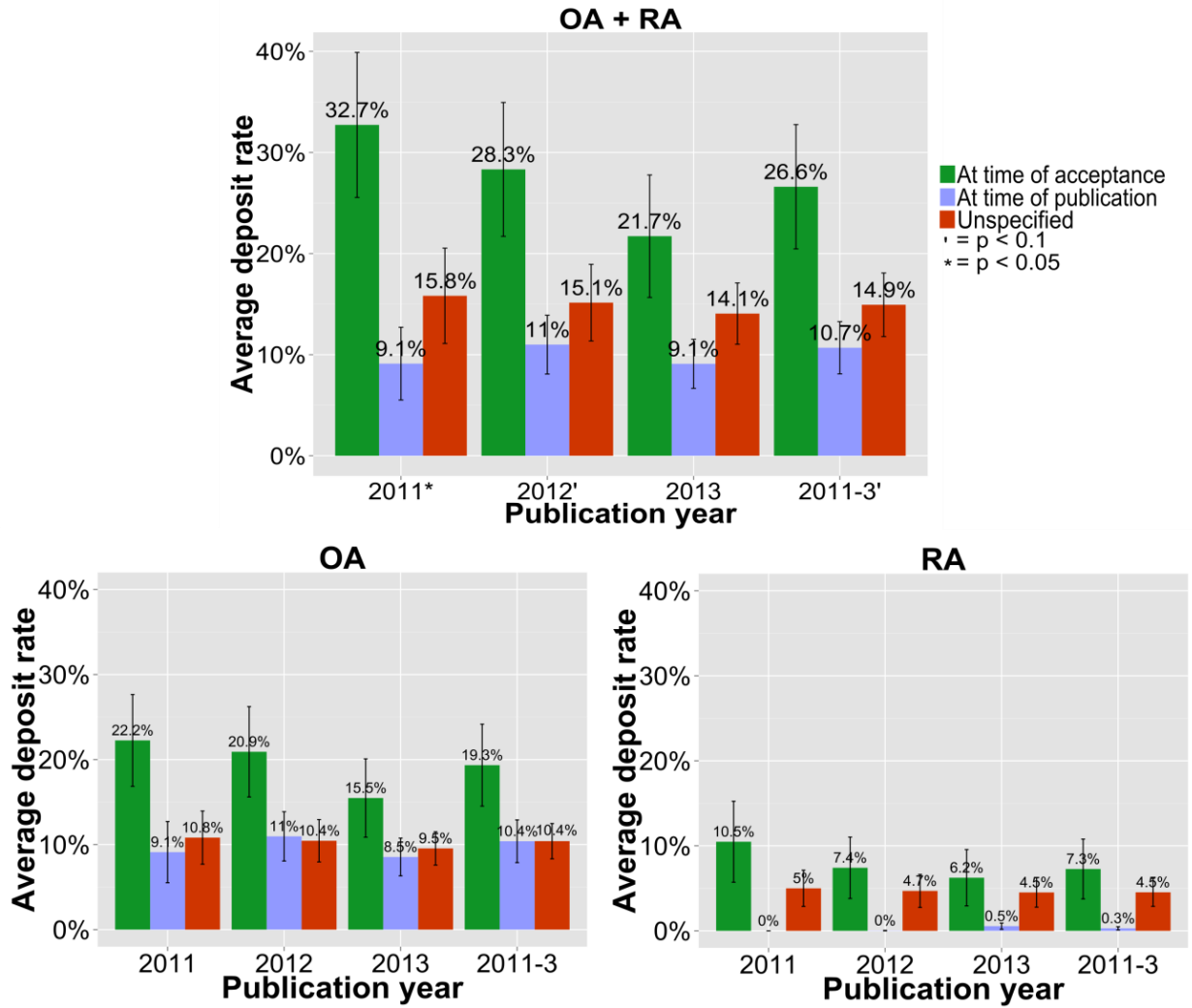

Figure 6: Average deposit rate when deposit is required at time of acceptance, at time of publication, or unspecified (Condition C4). For each of the three years, requiring deposit at time of acceptance consistently generates a higher deposit rate for RA + OA considered jointly (upper histogram) as well as for OA (left) and RA (right) considered separately (lower histogram).

*When* the deposit must be done is an especially important parameter in determining mandate effectiveness. Our interpretation is that requiring deposit at time of acceptance gives authors a much clearer and more specific time-marker than requiring deposit at time of publication. After submission, refereeing, revision and resubmission, successful authors receive a dated letter from the journal notifying them that their final draft has now been accepted for publication. That is the natural point in authors' workflow to deposit their final draft (Swan 2014): Authors know the acceptance date then, and have the final draft in hand. Publication date, in contrast, is uncertain: (1) authors don't know when the published version will appear: the delay ("publication lag") following acceptance can be quite long (sometimes months or even years); (2) the calendar date on the published issue may not correspond to the actual date it appeared; and (3) by the time the article is published the author may no longer have the final draft available for deposit. Requiring deposit at an unspecified date is even more vague, and probably closer to not requiring deposit at all.

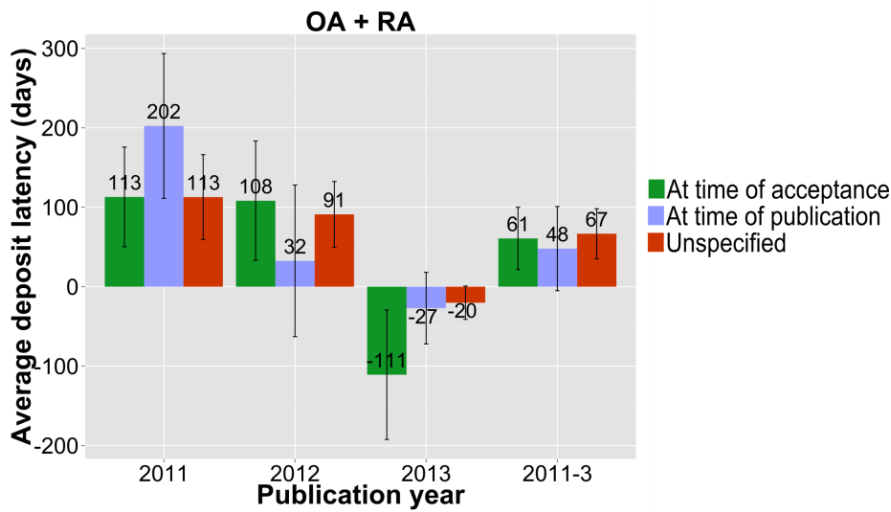

Figure 7: Average deposit latencies (delays relative to publication date = 0). There is no interpretable effect of deposit timing requirements (Condition C4) on average deposit latencies, either for RA + OA (shown), or for RA and OA separately.

One would expect the deposit timing requirement (Condition C4) to have an effect not only on deposit rate but also on deposit latency, i.e., the time at which the deposit is done; yet no interpretable pattern or statistical significance was observed across the three years for RA, OA or RA + OA (Figure 7). This may be because RA deposits are eventually converted to OA (as the OA embargo elapses). So the effect of deposit timing on deposit latency may be indirect, generating *more* (rather than earlier) *RA deposits* (which in any case occur earlier than OA deposits, and later become OA). Latency effects are harder to detect because our sample sizes for these measures are much smaller. The latency pattern is likely to become clearer as the database grows. We can already detect a significant effect of the internal-use condition (C7) on deposit latency (see below).

Which version must be deposited (the author's final draft or the publisher's version of record -- MELIBEA Condition C3) might also be expected to have an effect on deposit rate and/or latency, but it was not possible to make the comparison, because of our exclusion criteria: Before excluding the institutions that had fewer than 30 publications or Webometric rank beyond 5000, mandates requiring deposit of the author's version did have a significantly higher deposit rate than those requiring deposit of the publisher's version. But once the weaker institutions were excluded, almost all the remaining policies required the publisher's version, so no comparison was possible. This pattern, however, is consistent with the interpretation that requiring deposit immediately upon acceptance generates more deposits, and hence more deposits of the author's final draft. As there are more publisher restrictions on the later publisher's version than on the earlier author's version, and as many authors are reluctant to put their chances of getting published at (perceived) risk by challenging their publishers' restrictions, it is to be expected that the compliance rate for the author's version will be higher, and also that deposit will be done sooner, with papers initially deposited as RA and then made OA after the publisher OA embargo has elapsed.

## Effect of Internal-Use Condition (C7) on Deposit Rate and Latency

When it is stipulated that deposit is required "*For internal use*" (Condition C7), deposit rates are significantly higher for OA and RA deposits combined (Figure 8). The effect is larger and significant for RA deposits alone and smaller but in the same direction for OA deposits alone. With an internal-use requirement (usually for research performance evaluation), deposit latency is also significantly shorter for OA and RA combined, but this time the effect is larger and significant for OA deposits alone, smaller and not significant for RA deposits alone (Figure 9). (Note that especially for 2013 this measurement in April 2014 was probably too early for a reliable estimate because the average delay of about 175 days for mandates without an internal-use requirement had not yet been reached in April 2014. Similarly, the deposit rate for 2011 may be inflated because papers published in that year had the longest time to get deposited by April 2014.)

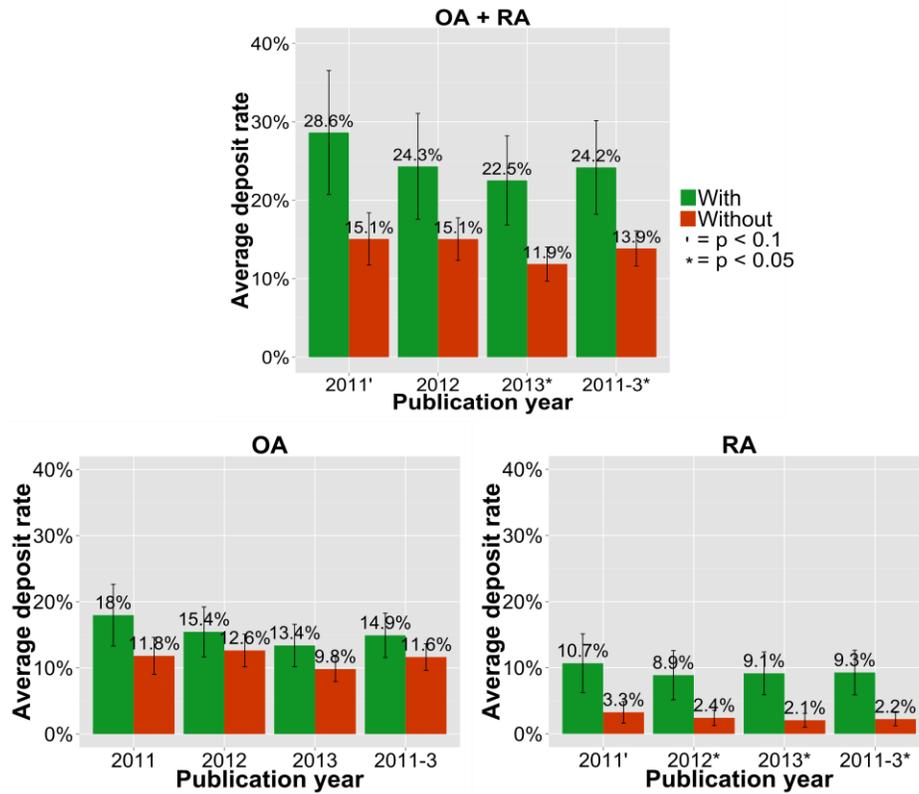

Figure 8: Average deposit rates, with and without internal use. For each of the three years, average OA + RA deposit rates are higher when the deposit is required for internal institutional use (e.g., research evaluation; Condition C7). The pattern is the same for OA alone (left) and RA alone (right), for each of the three years, but the deposit rate difference is bigger for RA.

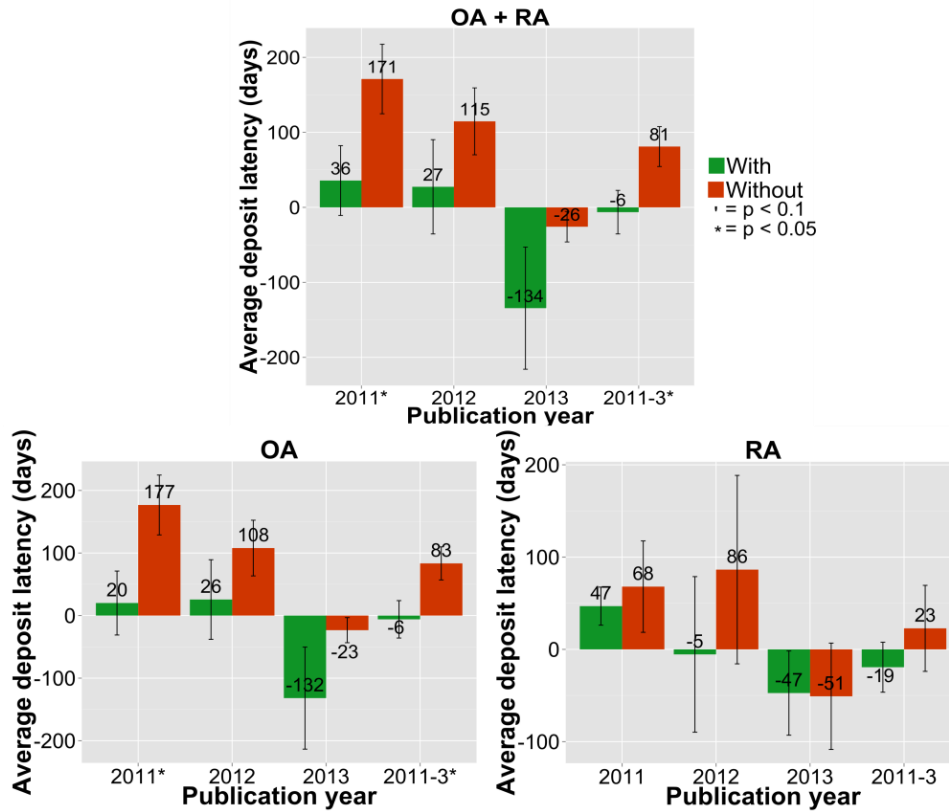

Figure 9: For each of the three years the average OA + RA deposit latencies are shorter when there is a requirement to deposit for internal use (Condition C7). (Higher is later, lower is earlier; 0 is the date of publication; latency is in days, relative to publication date). The pattern is the same for OA alone (left) and RA alone (right) for each of the three years, but the effect is bigger for OA.

The internal-use requirement pertains mostly to research performance evaluation, on which a researcher's rank and salary often depend. Hence it is predictable that researchers will be eager to make their papers available for this purpose by depositing them (Rentier & Thirion 2011). And since internal use only requires deposit, not OA, it is also predictable that this requirement will have a stronger effect on RA deposit than on OA deposit. What is a little surprising is that an internal-use requirement *accelerates* OA deposits more than RA deposits. This may be because our latency data were noisier and had more gaps (missing data) than our deposit data. It could also be because RA deposits occur earlier than OA deposits in any case (because of publisher OA embargoes), leaving less room for shortening their latency even further. Or it may be because the internal-use requirement (C7) has the effect of *reinforcing* the deposit-upon-acceptance requirement (C4) for OA deposits (those that have no publisher OA embargo with which the author wishes to comply): "You have to deposit immediately on acceptance anyway, for performance evaluation, so you may as well make it OA immediately too, rather than just RA."

## Effect of Opt-Out Condition (C2) on Deposit Rate and Latency

The last policy condition that proved significant concerned the right to waive or opt out of a requirement (Condition C2): We looked at the subset of mandates where the author was required both to deposit and to make the deposit OA, but was allowed to opt out of the OA requirement (only) (i.e., the deposit could be RA instead of OA). In particular, the effect concerned whether (i) unconditional opt-out from the OA requirement was allowed or (ii) the author had to negotiate each opt-out individually, on a conditional, case-by-case basis.

We found that an OA requirement allowing unconditional opt-out generates a higher deposit rate than an OA requirement allowing only a conditional opt-out (Figure 10). This result might seem paradoxical at first, because in a sense a "requirement" that allows an unconditional opt-out is not a requirement at all! So why would it generate a higher deposit rate than a requirement allowing only a conditional opt-out? We think the answer is in the component of the requirement from which no opt-out at all is allowed, namely, the deposit requirement itself: The internal-use requirement (C7) reinforces compliance with the immediate-deposit requirement (C4). In contrast, having to negotiate opt-out from the OA requirement case-by-case (conditional OA opt-out) inhibits compliance with the immediate-deposit requirement, whereas the possibility of unconditional opt-out from the OA requirement (C2) reinforces compliance with the immediate-deposit requirement.

(Suppose an author was reluctant about complying with an OA mandate. The mandate requires both (1) depositing and (2) making the deposit OA. If the author's reluctance was about having to make the deposit OA (i.e., 2) – because of legal worries, for example -- the author may not deposit at all (i.e., 1) if opting out of OA (2) could only be done through negotiations whose outcome was uncertain. But if authors whose ambivalence was because of (2) knew in advance that they could opt out of (2) unconditionally, without the need to negotiate, then they would no longer have any reason not to comply with (1), by depositing and leaving the deposit as RA instead of OA and relying instead on the repository's copy-request Button to provide access during the publisher embargo.)

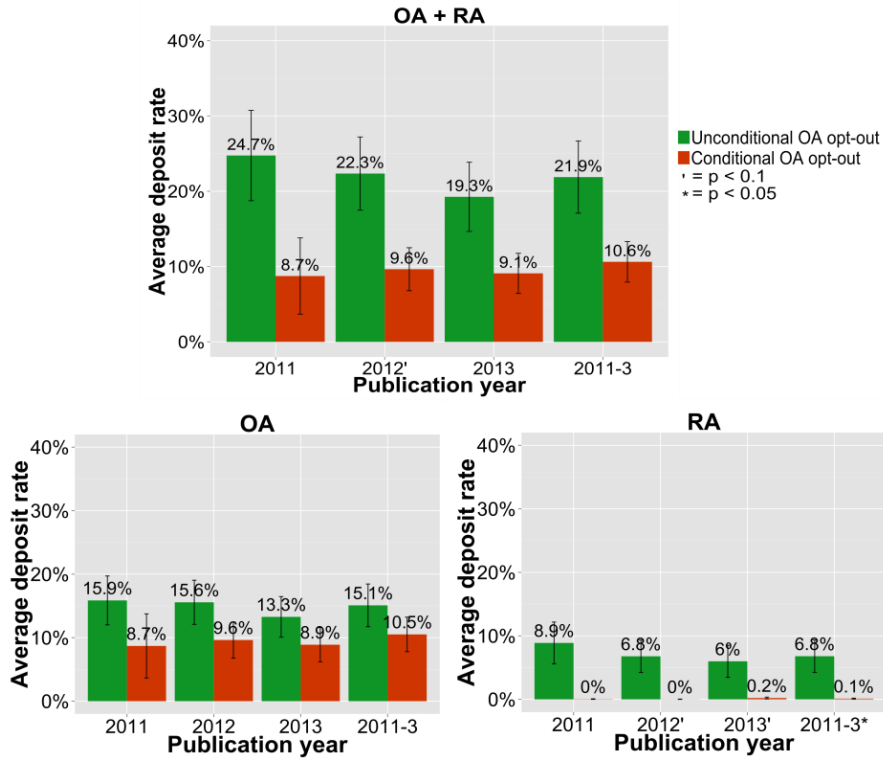

Figure 10: When deposit is required (no opt out) but authors can opt out of OA unconditionally (Condition C2), deposit rates are higher than when each individual OA opt-out has to be negotiated on a case-by-case basis. The effect is especially strong for RA (lower right); for each of the three years OA (lower left) and OA+RA (above) also show the same pattern as RA (above).

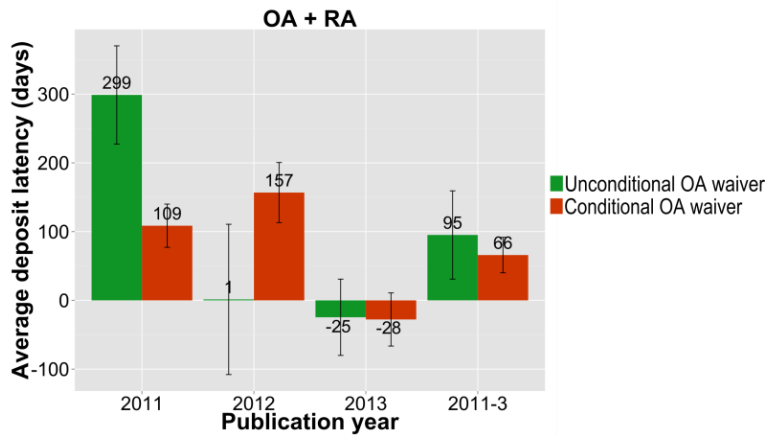

Figure 11: There is no consistent, interpretable effect of conditional vs. unconditional OA opt-out (Condition C2) on deposit latencies across the years. (Only OA + RA shown here)

All the data on deposit rates and deposit latencies can be thought of as measures of mandate *compliance rates*. In the case of the effect of allowing unconditional opt-out from the OA requirement, the biggest observed increase is in RA deposits, and our interpretation is that authors are more likely to comply with a deposit requirement if they can choose to deposit as RA rather than OA whenever they feel it is necessary, without the prospect of having to do individual case-by-case negotiation for a waiver of the OA requirement. Knowing that making the deposit OA is optional makes it more likely that an author will comply with the requirement to deposit-on-acceptance – or even to deposit at all.

## Power of Updated MELIBEA Score to Predict Deposit Rate

The original option-values and condition-weightings in the formula for the MELIBEA score were initialized largely on the basis of the recommendations of Peter Suber (2009) regarding mandate "strength" (including how many of the potential desiderata of OA they stipulate, such as Libre OA permissions and Gold OA funding). Based on their observed success rate for increasing and accelerating deposits, the initial MELIBEA values and weightings for mandate strength can now be updated as empirical evidence accrues so as to yield a formula with increased power to predict mandate *effectiveness* rather than just mandate strength. Using the current findings on the association of Conditions C2 (Opt-Out), C4 (Deposit Timing) and C7 (Internal Use), we have updated option-values and condition-weightings of MELIBEA to reflect their correlations with success: Higher values have been assigned to those options that our t-tests and ANOVAs showed to be more predictive of deposit rate and latency (Table 3). We have accordingly re-weighted the MELIBEA formula so as to reflect mandate effectiveness, assigning zero weight to the two irrelevant conditions (C1, Mandate/Request and C8, Theses) and equal weights to the rest (Figure 12).

$$0(C1) + 1(C2) + 1(C3) + 1(C4) + 1(C5) + 1(C6) + 1(C7) + 0(C8)$$

Figure 12: Updated values of the initial weights on the eight conditions (C1-C8), and the weights on the value options for each of these conditions, in MELIBEA's formula for mandate strength, as modified for predicting mandate effectiveness based on the findings of the present study. Condition C1 (Mandate or Request) was dropped because this study only considers mandates. Condition C8 (Theses) was dropped because this study considers only journal articles. The six other condition weights are set as one, so the substantive weighting updates were to the option values for each of the six non-zero conditions, as shown in Table 3, compared to Table 2.

| C1: Mandate or Request | | C5 : Embargo Length | |
|---|---|---|---|
| Value | | 0 | 12 months after publication |
| 0 | Request | 0.5 | 6 months after publication |
| 0 | Requirement | 0 | More than 12 months after publication |
| **C2 : Opt-out** | | 0 | Unspecified ('after period stipulated by the publisher') |
| 0 | No | **C6 : Copyright** | |
| **1** | **No deposit opt-out but unconditional OA opt-out** | 0 | Require blanket copyright reservation, no opt-out |
| **5** | **No deposit opt-out but conditional OA opt-out** | 0.5 | Allow opt-out from copyright reservation case by case |
| 0 | Deposit opt-out and unconditional OA opt-out | 0 | Retain copyright and license specific rights to publisher |
| **C3 : Version** | | 0.5 | Authors retain the non-exclusive rights of explotation to allow self-archiving |
| 2.5 | Deposit author's version | 1 | Authors should retain copyright whenever is possible |
| 0 | Deposit publisher's version | 0 | Any publishing or copyright agreements concerning articles have to comply with the OA policy |
| 0 | Unspecified | 0 | No copyright reservation |
| 0 | Unrefereed preprint | **C7 :  Internal use** | |
| **C4 : Deposit Timing** | | **5** | **Yes** |
| 0 | As soon as possible | **0** | **No** |
| **10** | **At time of acceptance** | **C8 : Theses** | |
| **0** | **At time of publication** | 0 | Yes |
| **0** | **Unspecified** | 0 | No |

Table 3: MELIBEA formula (Table 2) with values updated based on the findings of the present study. Higher values have been assigned to the options that t-tests and ANOVAs showed to be predictive of deposit rate and latency.

We make no claim that the new values and weightings are as yet optimal. A larger sample may well bring other conditions and their options into play as more mandates are adopted. In addition, latency effects will be clearer with a more complete and accurate sample than the present one. Nor are deposit rates and latencies the only criterion against which one might wish to validate mandate strength estimators. However, it is clear that our evidence-based updates of the initial values and weights already do increase the power of the MELIBEA Score to predict deposit rate (though not latency). The correlations in Figure 13 are all higher than those in Figure 5, and especially dramatically so for RA deposits.

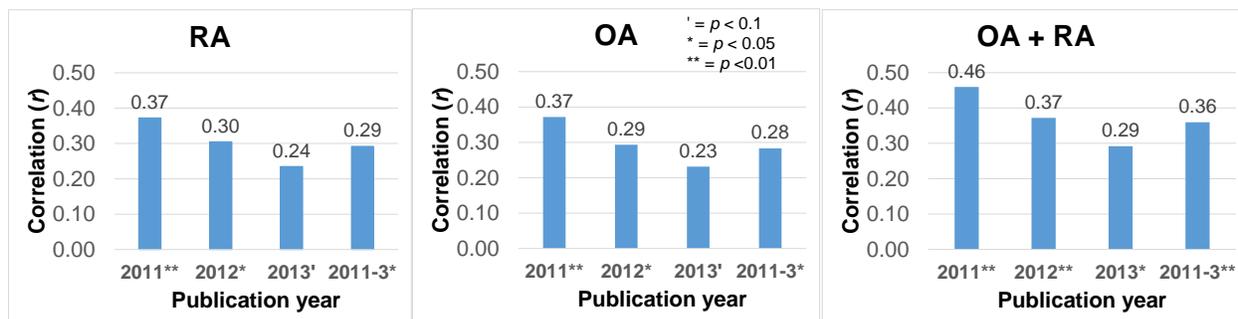

Figure 13: Correlation between updated MELIBEA mandate strength score and deposit rate, by publication year, for RA deposits, OA deposits and RA + OA deposits jointly, with values and weights updated to reflect the findings of this study. (Cf. Figure 5). (The correlations with latency remained inconsistent and uninterpretable.)

**Limitations**

This is an exploratory study in which we made *a priori* predictions as to which variables we expected to have an effect on deposit rates and latency, in which direction, and why. As only the three predicted variables had significant effects, we did not apply a Bonferroni correction for multiple *a posteriori* tests (though we did test many other potential variables). We take the replication of the direction of the effects (whether individually significant or insignificant) -- across the three independent time periods, and for RA + OA jointly as well as for RA and OA separately -- to further decrease the likelihood that the observed pattern of effects was due to chance*.*

(1) The data from multiple databases have limitations because they are incomplete, noisy and approximate. We have publication data from the top 12,000 journals, but we only have access to the full deposit and publication date of a portion of them. The latency data are especially noisy and incomplete.

(2) The short (3-year) time window for which we have deposit rate and latency data provides a narrowing picture of changes across time, especially for the most recent year, 2013 (as measured in April 2014). For the articles published earliest in the 3-year time window (2011) the deposit rates have stabilized and the latency averages are based on a long enough time-base (2.5 years), whereas the deposit rates for articles published in the second half of 2013 are based on a time window of less than a year.

(3) MELIBEA's classification and weighting of OA mandates concerned mandate "strength" rather than mandate effectiveness: its initial option-values and condition-weights were hence not intended, validated or optimized for predicting effectiveness. We have now updated them empirically to predict deposit rate based on the data available to date. As the number and age of the mandates increases and the ranking and classification system is updated and optimized further, the data should become less noisy and variable and the effectiveness prediction can be further optimized. Moreover, if new mandates adopt the conditions that have already emerged as most important from the present analysis, both the rate and speed of deposit are likely be enhanced.

(4) As stated, the analysis is based only _on (non-OA)_ publications indexed by WoS, and the criterion for inclusion was at least 30 WoS-indexed publications within the 3-year window. It is very possible and likely that all institutions also had non-WoS-indexed publications in that interval, but only WOS-indexed articles were used in order to ensure that we were comparing like with like. Future studies will look at total publication output, but for OA purposes it is fair to say that WoS-indexed publications are the first priority test case. In particular, not only are those the publications that most users need most, but, even more important, those are the publications that are most likely to be embargoed by their publishers, thereby constraining deposit rates and mandate effectiveness

## Summary and Conclusions

Our first finding was that the higher an institution's MELIBEA score for OA mandate strength, the higher was the rate of deposit in its institutional repository for each year, but the correlation was small, and was significant only for OA + RA deposits jointly and for OA deposits alone, but not for RA deposits alone. There was also no correlation with latency.

In an effort to increase the predictive power of the MELIBEA score for mandate effectiveness rather than just strength, and to determine the effects of its conditions individually, we analyzed three of the eight conditions separately to test their effects on deposit rate and latency. (Two of the other MELIBEA conditions were irrelevant because they pertained to non-mandates or to theses, respectively, and for the remaining three the sample was too small to test them separately.) The three conditions we examined were (C4) _when the author had to do the deposit_ (upon acceptance, upon publication, or unspecified; (C7) _whether the deposit was required for internal use_ (such as performance evaluation) and (C2) _whether the author could opt out of the requirement to make the deposit OA_ unconditionally or only conditionally, on a case-by-case basis.

For the deposit-timing condition (C4), the requirement to deposit immediately upon acceptance generated significantly higher deposit rates for OA + RA deposits combined. The pattern was also the same for OA and RA separately, and for each individual year. For deposit latency there was no consistent or significant effect.

For the internal-use condition (C7), requiring deposit for internal use generated significantly higher deposit rates for OA + RA deposits combined. The pattern was also the same for RA and OA separately, and for each individual year, but the rate-increase was larger for RA than for OA. The internal-use requirement had an effect on deposit latency as well: Deposit is done earlier when it is required for internal use (same pattern for RA + OA, RA, OA, and for each year), but the latency-shortening effect is stronger for OA than for RA.

Our interpretation for C7 is that when the deposit is required for internal use (which mostly means research performance evaluation), more authors who may feel inhibited by a publisher OA embargo from depositing OA go ahead and deposit as RA rather than not depositing at all or waiting for the end of the publisher OA embargo to deposit. This is what increases the rate of RA deposit. For the OA deposits, which tend to be done later than RA deposits (because of publisher OA embargoes), the effect of their being needed for internal use is to reduce their latency (i.e., speed them up).

For the opt-out condition (C2), allowing unconditional opt-out from the OA requirement (but not the deposit requirement) generated significantly higher deposit rates for OA + RA deposits combined. The pattern was the same for RA and OA separately, and for each individual year, but the effect was especially marked for RA. (There was no effect on latency.) Our interpretation for C2 is that authors are more likely to comply with a deposit requirement if they know they can choose unconditionally to deposit as RA rather than OA.

Using our findings on the effects of these three conditions, we updated the initial option-values and condition-weightings of the original MELIBEA formula with the result that its correlation with deposit rate increased overall, and especially for RA deposits. There is still no correlation with deposit latency, but as RA deposits eventually become OA deposits, increases in RA may reflect earlier deposit of papers that might otherwise only have been deposited later, as OA. (Note that RA deposits are indirectly accessible even during the OA embargo through individual user requests to authors via the repository's copy-request Button; Sale et al 2014.)

Our updating of a subset of the MELIBEA values and weights shows that in combination, three (at least) of the MELIBEA parameters can already predict deposit rate – and hence reflect mandate effectiveness – more closely than the original MELIBEA overall score, which was designed to reflect mandate strength. This is just an illustration of how further research can be used to keep optimizing the predictive power of estimators of mandate effectiveness.

Practically speaking, it also emerges from this analysis that the deposit-on-acceptance requirement, internal-use requirement, and the possibility of unconditionally opting out of making the deposit OA are each very important factors in the effectiveness of an OA mandate in generating greater author compliance and hence more deposits and more OA, sooner. As the rate of adoption of OA policies is now growing too, these findings make it possible for OA policy-making to become more evidence-based and more effective by taking these findings into account in designing OA mandates.[2]

---

[2] There are many other factors that might increase deposit rates, with or without mandates. Having a CRIS is one of them. Having IRstats that give authors feedback on

This study was based only on institutional mandates rather than funder mandates, because it is far easier to identify total journal article output for an institution's researchers (using the WoS institutional tag) than for a funder's grantees (which spans many institutions), but there is no reason to think the conclusions do not apply to both kinds of mandates.

**References**

Archambault, É. (2013). The Tipping Point: Open Access Comes of Age. In ISSI 2013 *Proceedings of 14th International Society of Scientometrics and Informetrics Conference* (Vol. 1, pp. 1165-1680). http://users.ecs.soton.ac.uk/harnad/Temp/ISSI-ARchambeault.pdf

Berners-Lee, T. (1989). *Information management: A proposal.* http://www.w3.org/History/1989/proposal.html

Björk B.C., Welling, P., Laakso, M., Majlender, P., Hedlund, T., & Guðni, G. (2010). Open Access to the scientific journal literature: Situation 2009. *PLOS ONE*, 5(6), e11273. http://dx.plos.org/10.1371/journal.pone.0011273

Björk, B. C., Laakso, M., Welling, P., & Paetau, P. (2014). Anatomy of green open access. *Journal of the Association for Information Science and Technology*, 65(2), 237-250. http://openaccesspublishing.org/apc8/Personal%20VersionGreenOa.pdf

BOAI (2002) *Budapest Open Access Initiative.* http://www.budapestopenaccessinitiative.org

Carr, Leslie and Harnad, Stevan (2005) Keystroke Economy: A Study of the Time and Effort Involved in Self-Archiving. *University of Southampton Working Paper*. http://eprints.soton.ac.uk/260688/

Carr, L., Harnad, S., & Swan, A. (2007). A Longitudinal Study of the Practice of Self-Archiving. *University of Southampton Working Paper*. http://eprints.soton.ac.uk/263906/

Chen, X. (2014). Open Access in 2013: Reaching the 50% Milestone. *Serials Review*, 40(1), 21-27.

Edgington, E., & Onghena, P. (2007). *Randomization tests.* CRC Press.

ETH Zurich FAQ (2015) *Legal questions on self-archiving in ETH E-Collection* https://www.library.ethz.ch/en/ms/Open-Access-at-ETH-Zurich/Legal-aspects-of-open-access-publishing/Legal-questions-on-self-archiving-in-ETH-E-Collection

Gargouri, Y, Larivière, V & Harnad, S (2013) Ten-year Analysis of University of Minho Green OA Self-Archiving Mandate In E Rodrigues, A Swan & AA Baptista, Eds. *Uma* the downloads of their articles is another. Having automatic SWORD-based import/export to and from other repositories is another. Providing librarian-facilitated proxy deposit is another.

*Década  de Acesso Aberto e na UMinho no Mundo*. U Minho
http://eprints.soton.ac.uk/358882/

Gargouri, Y., Larivière, V., Gingras, Y., & Harnad, S. (2012). Green and gold Open Access percentages and growth, by discipline. In É. Archambault, Y. Gingras, & V. Larivière (Eds.), *Proceedings of 17th International Conference on Science and Technology Indicators*. Montréal, Canada
http://eprints.soton.ac.uk/340294/1/stiGargouri.pdfhttp://eprints.soton.ac.uk/340294/1/stiGargouri.pdf

Gargouri, Y., Hajjem, C., Larivière, V., Gingras, Y., Carr, L., Brody, T., & Harnad, S. (2010). Self-selected or mandated, open access increases citation impact for higher quality research. *PloS one*, *5*(10). http://dx.plos.org/10.1371/journal.pone0013636

Harnad, S. (2006) Opening Access by Overcoming Zeno's Paralysis, in Jacobs, N., Eds. Open Access: *Key Strategic, Technical and Economic Aspects*. Chandos.
http://eprints.ecs.soton.ac.uk/12094/

Harnad, S., Brody, T., Vallieres, F., Carr, L., Hitchcock, S., Gingras, Y, Oppenheim, C., Stamerjohanns, H., & Hilf, E. (2004) The green and the gold roads to Open Access. Nature Web Focus. http://www.nature.com/nature/focus/accessdebate/21.html

Harvard University Library (2012) Major Periodical Subscriptions Cannot Be Sustained. *Faculty Advisory Council Memorandum on Journal Pricing*
http://isites.harvard.edu/icb/icb.do?keyword=k77982&tabgroupid=icb.tabgroup143448

Hitchcock, S. (2013) *The effect of open access and downloads ('hits') on citation impact: a bibliography of studies*. http://opcit.eprints.org/oacitation-biblio.html
http://opcit.eprints.org/oacitation-biblio.html

Khabsa, M., & Giles, C. L. (2014). The Number of Scholarly Documents on the Public Web. *PloS ONE*, 9(5), e93949.
http://www.plosone.org/article/info:doi/10.1371/journal.pone.0093949%22%20%5Cl%20%22pone-0093949-g003

MELIBEA (2015) Directory and estimator policies for open access to scientific production. http://www.accesoabierto.net/politicas http://www.accesoabierto.net/politicas

Miller, FP, Vandome, AF, McBrewster J (2010) *Serials Crisis*. VDM Publishing

Okerson, A., & O'Donnell, J. J. (Eds.). (1995). Scholarly journals at the crossroads: a subversive proposal for electronic publishing. *Association of Research Libraries*.
http://catalog.hathitrust.org/Record/003013520

Pinfield, S. (2001). How do physicists use an e-print archive? Implications for institutional e-print services. *D-Lib Magazine*, 7(12) http://eprints.nottingham.ac.uk/51/

Rentier, B., & Thirion, P. (2011). The Liège ORBi model: Mandatory policy without rights retention but linked to assessment processes. Berlin 9 Prconference Workshop, November 2011, http://orbi.ulg.ac.be/jspui/bitstream/2268/102031/1/Rentier-WashDC-2011.pdf

ROAR (2015) Registry of Open Access Repositories. http://roar.eprints.org/information.html

ROARMAP (2015) Registry of Open Access Repositories Mandatory Archiving Policies. http://roarmap.eprints.org/

Sale, A., Couture, M., Rodrigues, E., Carr, L. and Harnad, S. (2014) Open Access Mandates and the "Fair Dealing" Button. In: *Dynamic Fair Dealing: Creating Canadian Culture Online* (Rosemary J. Coombe & Darren Wershler, Eds.) http://eprints.ecs.soton.ac.uk/18511/

Self-Archiving FAQ (2015) http://www.eprints.org/openaccess/self-faq/

SHERPA/Romeo (2015) *Publisher copyright policies & self-archiving*. http://www.sherpa.ac.uk/romeo/

Spezi, V., Fry, J., Creaser, C., Probets, S., & White, S. (2013). Researchers' green open access practice: a cross-disciplinary analysis. *Journal of Documentation*, 69(3), 334-359. https://dspace.lboro.ac.uk/dspace-jspui/bitstream/2134/12324/3/Green_OA_practice_and_Disciplinary_differences_revised_version_for%20repository.pdf

Suber, P. (2009) Open access policy options for funding agencies and universities. *SPARC Open Access Newsletter* 130 February 2009 http://legacy.earlham.edu/~peters/fos/newsletter/02-02-09.htm

Suber, P. (2008a) Gratis and Libre Open Access. *SPARC Open Access Newsletter 124* August 2008 http://www.sparc.arl.org/resource/gratis-and-libre-open-access

Suber, P. (2008b) An open access mandate for the NIH. *SPARC Open Access Newsletter 117*, January 2008. http://nrs.harvard.edu/urn-3:HUL.InstRepos:4322583

Swan, A. (2014) HEFCE announces Open Access policy for the next REF in the UK. *LSE Impact of Social Sciences blog*, 1 April 2014 http://blogs.lse.ac.uk/impactofsocialsciences/2014/04/01/hefce-open-access-ref-gamechanger/

Swan, A., Gargouri, Y., Hunt, M & Harnad, S (2015) Open Access Policy: Numbers, Analysis, Effectiveness. *Pasteur4OA Workpackage 3 Report*. http://eprints.soton.ac.uk/375854/

Swan, A., & Brown, S. (2005) Open access self-archiving: An author study. *JISC Report.*

Taylor & Francis (2014) Open Access Survey: examining the changing views of Taylor & Francis authors http://www.tandfonline.com/page/openaccess/opensurvey/2014

Thomson-Reuters/ISI Web of Science. (2015). http://wokinfo.com

Wallace, J. M. (2011). PEER: green open access–insight and evidence. *Learned Publishing*, 24(4), 267-277. http://www.peerproject.eu/fileadmin/media/ppt_about_peer/JWallace_LearnedPubl_Oct 2011.pdf

Webometrics (2015). Ranking web of universities. http://www.webometrics.info/en/Methodology

Xia, J., Gilchrist, S. B., Smith, N. X., Kingery, J. A., Radecki, J. R., Wilhelm, M. L., ... & Mahn, A. J. (2012). A review of open access self-archiving mandate policies. *portal: Libraries and the Academy*, *12*(1), 85-102. http://www.press.jhu.edu/journals/portal_libraries_and_the_academy/portal_pre_ print/archive/articles/12.1xia.pdf